\documentclass[]{aa}
\usepackage{graphicx}
\usepackage{txfonts}
\begin{document}
\title{Magnetic field confinement by meridional flow and the
       solar tachocline}
\titlerunning{Magnetic solar tachocline}
   \author{L.\,L.~Kitchatinov\inst{1,2} \and G.~R\"udiger\inst{1}
          }
   \offprints{G.~R\"udiger}
   \institute{Astrophysikalisches Institut Potsdam, An der Sternwarte 16,
              D-14482, Potsdam, Germany \\
              \email{gruediger@aip.de; lkitchatinov@aip.de}
         \and
             Institute for Solar-Terrestrial Physics, PO Box
             4026, Irkutsk 664033, Russian Federation\\
             \email{kit@iszf.irk.ru}
             }
   \date{Received ; accepted }
   \abstract{We show that the MHD theory that explains the solar tachocline by
   an effect of the magnetic field can work with the decay modes of a fossil
   field in the solar interior if the meridional flow of the convection
   zone penetrates slightly the radiative zone beneath. An equatorward
   flow of about 10 m/s penetrating to a maximum depth of 1000 km below the
   convection zone is able to
   generate almost horizontal field lines in the tachocline region so that the
   internal field is almost totally confined to the radiative zone. The theory of
   differential solar rotation indeed provides meridional flows of about 10 m/s
   and a penetration depth of {\lower.4ex\hbox{$\;\buildrel <\over{\scriptstyle\sim}\;$}}
   1000~km for viscosity values that are characteristic of a {\em stable}
   tachocline.
   }
\maketitle
\section{Introduction}
While rotation of the solar radiative core is almost uniform, the
convection zone rotates differentially (Wilson et al.
\cite{WBL97}; Kosovichev et al. \cite{KSS97}; Shou et al.
\cite{Sea98}). The transition from inhomogeneous to rigid rotation
occurs in a thin almost spherical layer named \lq tachocline',
whose parameters are well known from helioseismology. The
tachocline thickness is estimated to be
\lower.4ex\hbox{$\;\buildrel <\over{\scriptstyle\sim}\;$}$0.05
R_\odot$, its midpoint radius is $\left(0.692 \pm 0.005\right)
R_\odot$, and it is slightly prolate in shape (Kosovichev
\cite{K96}; Antia et al. \cite{ABC98}; Charbonneau et al.
\cite{Cea99}). The tachocline lies totally, or at least mostly,
beneath the base of convection zone at $R_\mathrm{in} = 0.713
R_\odot$ (Christensen-Dalsgaard et al. \cite{CDea91}; Basu \&
Antia \cite{BA97}), at the very top of the radiative core.

The existence of tachocline is evidence of a link between low and
high latitudes that should be present just beneath the convection
zone in order to suppress the latitudinal differential rotation.
Similarly, uniform rotation in the radius beneath the tachocline
indicates that a coupling in radius should be present there.
Otherwise, rotational braking of the Sun would leave a rapidly
rotating core. Among several theoretical concepts suggested for
the solar tachocline, only one that explains it by the effect of a
weak internal magnetic field can provide both links
simultaneously (Charbonneau \& MacGregor \cite{CM93}; MacGregor
\& Charbonneau \cite{MC99}). The uniform rotation of the core was
even considered as evidence of an internal field. Another
argument by Cowling (\cite{C45}) is that the time of resistive
decay for the hypothetical internal field exceeds the solar age.

It has been confirmed by numerical modeling that even a weak
($\sim 10^{-3}$G) poloidal field can produce the tachocline
(R\"udiger \& Kitchatinov \cite{RK97}; MacGregor \& Charbonneau
\cite{MC99}). However, this is the case only if the field
geometry satisfies certain constraint. Ferraro's (\cite{F37}) law
on the constancy of angular velocity along the field lines
requires the field to be almost horizontal inside the tachocline
region (except, perhaps, for the mid latitudes where angular
velocity at the base of convection zone equals that of the bulk
of the core).

Why the field should be confined inside the radiative core has
remained, however, uncertain. The main idea of the present paper
is that the required field geometry may result from the influence
of the meridional flow penetrating the radiative core from the
convection zone. If the electric conductivity in the radiative
core is much higher than in the convection zone, then (and only
then) the flow has massive electrodynamical consequences, and the
poloidal magnetic field becomes parallel to the meridional flow
(cf. Mestel \cite{M99}).

The flow -- produced in the convection zone by centrifugal and
barocline forcing -- only slightly penetrates the radiative core
(Gilman \& Miesch 2004; Kitchatinov \& R\"udiger \cite{KR05}).
Its induction can thus be included via the boundary condition. In
this paper we formulate the boundary condition for a poloidal
field on the top of the radiative core with account for the
shallow penetration of meridional flow. With such a modified
boundary condition, the modes of the internal poloidal field with
the longest decay times are computed. The long-living modes show
the required confined structure. The thus defined internal field
is then used in tachocline computations to confirm that a slender
tachocline can indeed be found.
\section{Penetration}\label{estimations}
\begin{figure}[htb]
   \centering
   \includegraphics[width=4.12cm, height =4.3cm]{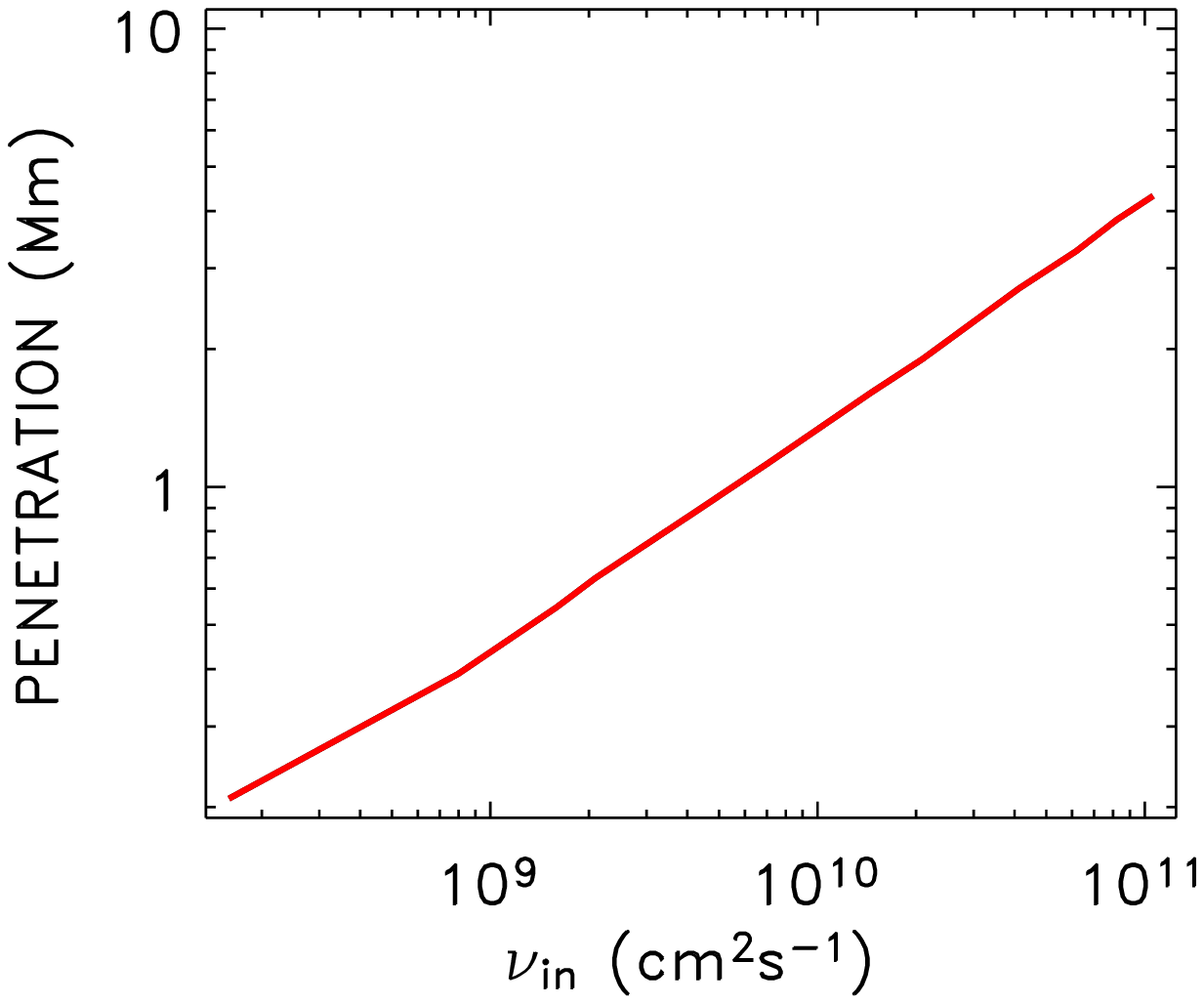}
   \hspace{0.2cm}
   \includegraphics[width=4cm, height =4.3cm]{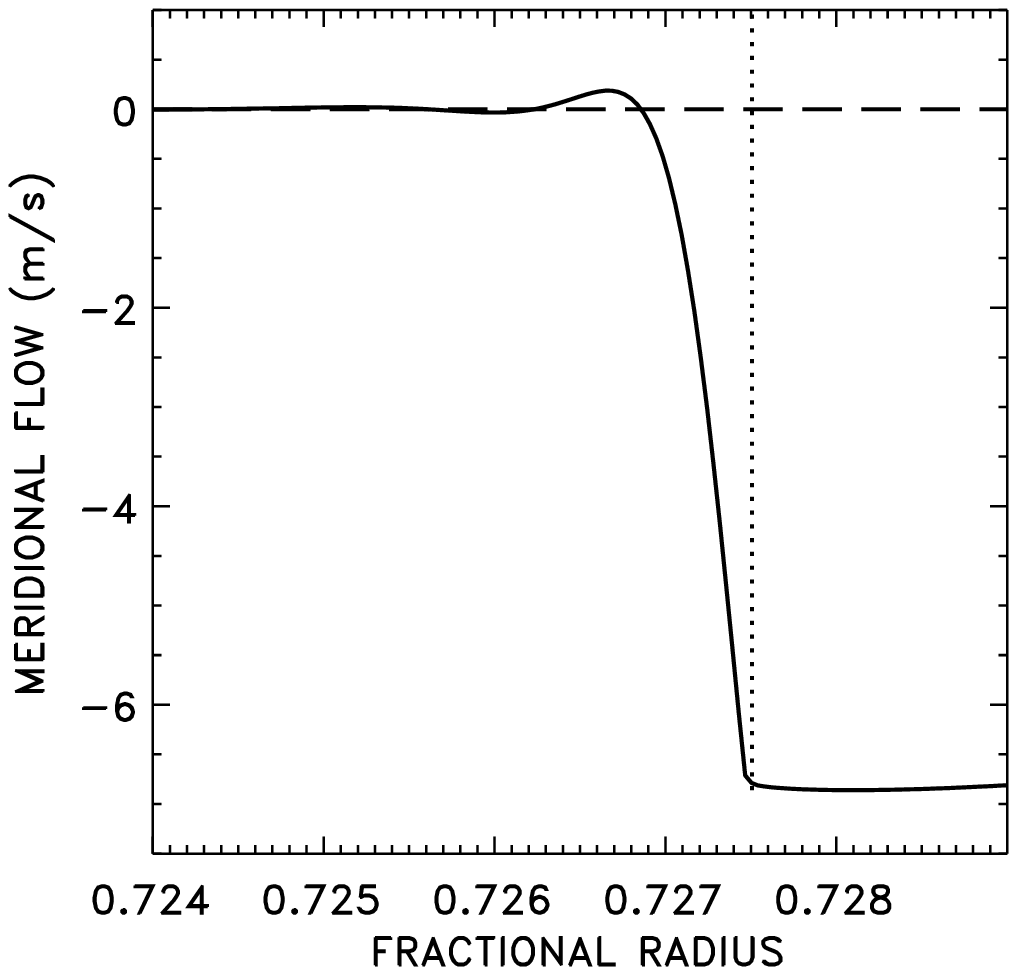}
   \caption{Left: penetration depth as a function of the core viscosity.
            Right: meridional velocity profile in the penetration
        region at  $45^\circ$ latitude for
        $\nu_\mathrm{in} = 1.1\cdot 10^9$\,cm$^2$s$^{-1}$.
        Negative values mean equatorward flow. The
        vertical dotted line marks the bottom of the
        convection zone.
              }
   \label{f1}
\end{figure}

The penetration of the meridional flow beneath the solar
convection zone was computed by Kitchatinov \& R\"udiger
(\cite{KR05}). The main results are shown in Fig.~\ref{f1}. The
meridional velocity in the penetration region shows reversals with
increasing depth, where the flow amplitude reduces largely after
each reversal. The penetration depth  $D_\mathrm{pen}$ in
Fig.~\ref{f1} has been defined as the distance from the base of
the convection zone to the location of the first reversal.

The penetration results from viscous drag imposed by the
meridional flow at the base of the convection zone on the fluid
beneath, and this is opposed by the Coriolis force. The Ekman
balance results in the estimation
$D_\mathrm{pen}\sim\sqrt{\nu_\mathrm{in}/\Omega}$ (Gilman \&
Miesch \cite{GM04},   $\nu_\mathrm{in}$ is the viscosity in the
penetration region, $\Omega$ its basic angular velocity). The
plot  of Fig.~\ref{f1} (left) is approximated well by
$D_\mathrm{pen} \simeq 2.3\sqrt{{\nu_\mathrm{in}}/\Omega}$ or,
equivalently, by
\begin{equation}
   D_\mathrm{pen} = 1.4\sqrt{\nu_\mathrm{in}}\cdot 10^3\ \mathrm{cm}
   \label{1}
\end{equation}
for solar parameters. Thus, the penetration distance is about
100~m for microscopic viscosity and remains smaller than 1~Mm for
any reasonable value of eddy viscosity. The penetration depth is
small even compared to the tachocline thickness,
$D_\mathrm{pen}\ll D_\mathrm{tach}$.

This shallow penetration may still strongly influence the geometry
of an internal poloidal field. The penetration makes a shear flow
on the top of the radiative core. The time $\tau_\mathrm{s} \simeq
D_\mathrm{p}/u^\mathrm{m}$ of latitudinal field shearing from the
radial one is small compared to the time $\tau_\mathrm{diff} =
D^2_\mathrm{pen}/\eta_\mathrm{in}$ of diffusive escape of the
field from the penetration region. The time ratio,
$\tau_\mathrm{diff}/\tau_\mathrm{s}$, gives the magnetic Reynolds
number,
\begin{equation}
    \mathrm{Rm} = \frac{D_\mathrm{pen} u^\mathrm{m}}{\eta_\mathrm{in}}
    \simeq
    \frac{\mathrm{Pm}}{\sqrt{\nu_\mathrm{in}}} 10^6 ,
    \label{2}
\end{equation}
where Pm is the magnetic Prandtl number. The value $u^\mathrm{m}
\simeq 10$\,m\,s$^{-1}$ is used for the meridional velocity. With
microscopic diffusivities, the number is large, $\mathrm{Rm}\sim
10^3$, and it remains higher than this value with eddy
diffusivities up to $10^6$\,cm$^2$\,s$^{-1}$. It sinks to unity
for eddy diffusivities of about $10^{12}$\,cm$^2$\,s$^{-1}$, so
that only in this case does the influence of a penetrating
meridional flow on the magnetic field geometry turn weak.

The shearing of the poloidal field is probably the only process
where this penetration is important. The advection time,
$\tau_\mathrm{adv} = R_\mathrm{in}/u^\mathrm{m}$, is much longer
than the diffusion time $\tau_\mathrm{diff}$,
\begin{equation}
    \frac{\tau_\mathrm{diff}}{\tau_\mathrm{adv}} =
    \frac{D_\mathrm{pen}}{R_\mathrm{in}}\ \  \mathrm{Rm} \simeq
    0.03\cdot \mathrm{Pm} \ll 1,
    \label{3}
\end{equation}
independent of whether  microscopic or eddy diffusion applies.
R\"udiger et al. (\cite{RKA}) argued that the tachocline region
should be stable in the hydrodynamic regime so that microscopic
diffusivities have to be used. In any case the diffusion time  is
so short that the penetration layer is not dynamo-relevant
(R\"udiger et al. 2005).
\section{The model}
Axial symmetry for both magnetic and velocity fields is assumed.
In spherical coordinates $\left( r,\theta,\phi\right)$, the fields
can be written as
\begin{equation}
  {\vec u} = \left(\frac{1}{\rho r^2\sin\theta}
  \frac{\partial\psi}{\partial\theta},
  \frac{-1}{\rho r\sin\theta}\frac{\partial\psi}{\partial r},
  r\sin\theta\ \Omega\right) ,
  \label{4}
\end{equation}
\begin{equation}
  {\vec B} = \left(\frac{1}{r^2\sin\theta}
  \frac{\partial A}{\partial\theta},
  \frac{-1}{r\sin\theta}\frac{\partial A}{\partial r},\
  B\ \right) ,
  \label{5}
\end{equation}
in terms of the stream function ($\psi$) of the meridional flow
and the   potential  ($A$) of the poloidal field. Here $\Omega$ is
the angular velocity and $B$ the toroidal field.

The poloidal field equation,
\begin{equation}
   \frac{\partial A}{\partial t} =
   -\frac{u_\theta}{r}\frac{\partial A}{\partial\theta}
   - u_r \frac{\partial A}{\partial r}
   + \eta\frac{\partial^2 A}{\partial r^2}
   + \frac{\eta}{r^2}\sin\theta\frac{\partial}{\partial\theta}
   \left(\frac{1}{\sin\theta}\frac{\partial A}{\partial\theta}\right) ,
   \label{6}
\end{equation}
decouples from other equations of the problem. As the eddy
diffusivity in the convection zone is large compared to
$\eta_\mathrm{in}$, the vacuum boundary condition can be applied
to the top of the core. The (nonlocal) condition,
\begin{equation}
   \frac{\partial A}{\partial r} =
   \left(\frac{\partial A}{\partial r}\right)_\mathrm{vac}\ \ \ {\rm at}\ \ \
   r = R_\mathrm{in},
   \label{7}
\end{equation}
is usually expressed in terms of the Legendre polynomial expansion
\begin{eqnarray}
   A\left( r,\theta\right) &=& \sin\theta\sum\limits_{n=1}^\infty
   A_n\left( r\right) P_n^1\left(\cos\theta\right) ,
   \nonumber \\
   \left(\frac{\partial A}{\partial r}\right)_\mathrm{vac} &=&
   - \frac{\sin\theta}{r} \sum\limits_{n=1}^\infty
   n A_n\left( r\right) P_n^1\left(\cos\theta\right) .
   \label{8}
\end{eqnarray}

The internal magnetic field  can be defined by solving
Eq.~(\ref{6}) with the boundary condition (\ref{7}). The
meridional flow is, however, only present in the very thin
penetration layer (Fig.~\ref{f1}). Instead of resolving this layer
explicitly, we integrate Eq.~(\ref{6}) across the layer to obtain
a new boundary condition for the poloidal field for the base of
the penetration layer that is still above the tachocline. The
effect of penetration is then included in the reformulated
boundary condition.
\subsection{Poloidal field boundary condition}
Inside the penetration layer, the advection time ($\sim 10$\,yr)
is short compared to the latitudinal diffusion time,
$R^2_\mathrm{in}/\eta_\mathrm{in} \sim 10^9$\,yr. The left part of
Eq.~(\ref{6}) and its last term can thus be neglected compared to
the first term on the right.

Also, the radial velocity in the penetration layer is small
compared to the latitudinal velocity, $u_r \sim u_\theta
D_\mathrm{pen}/R_\mathrm{in}$ (Gilman \& Miesch \cite{GM04}). By
neglecting these small terms the equation for the penetration
region reads
\begin{equation}
    \frac{\partial^2 A}{\partial r^2} +
    \frac{1}{\eta\rho R_\mathrm{in}^2\sin\theta}
    \frac{\partial\psi}{\partial r}
    \frac{\partial A}{\partial\theta} = 0 ,
    \label{9}
\end{equation}
where $u_\theta$ is expressed in terms of the stream function of
Eq.\,(\ref{4}). The stream function varies from zero at the base
of the layer to a finite value at its top. Divergence-free of
magnetic field requires a variation in the radial field, $B_r =
(\partial A/\partial\theta )/(R^2_\mathrm{in}\sin\theta )$,
across the layer be small. Integration of Eq.\,(\ref{9}) across
the layer then yields
\begin{equation}
    \frac{\partial A}{\partial r} -
    \frac{\psi (R_{\rm in},\theta )}{\eta_\mathrm{in}\rho R_\mathrm{in}^2\sin\theta}
    \frac{\partial A}{\partial\theta} =
    \left(\frac{\partial A}{\partial r}\right)_\mathrm{vac},
    \label{10}
\end{equation}
where $\psi(R_{\rm in},\theta )$ is the stream function profile
at the base  of the convection zone. The stream function scales as
\begin{equation}
    \psi (R_{\rm in},\theta ) = u^\mathrm{m}\rho R_\mathrm{in}
    D_\mathrm{pen}\hat\psi (\theta ),
    \label{11}
\end{equation}
where $\hat\psi$ is a dimensionless function of order one.
Substitution of (\ref{11}) into (\ref{10}) leads to the boundary
condition sought for the poloidal field,
\begin{equation}
   R_\mathrm{in}\frac{\partial A}{\partial r} -
   \mathrm{Rm}
   \frac{\hat\psi\left(\theta\right)}{\sin\theta}
   \frac{\partial A}{\partial\theta} =
   R_\mathrm{in}\left(\frac{\partial A}{\partial r}\right)_\mathrm{vac},
   \label{12}
\end{equation}
which accounts for the shallow penetration of the meridional flow
from the convection zone into the radiative interior.
\subsection{Model equations}
Any meridional flow in the bulk of radiative core is neglected.
The characteristic time of the Eddington-Sweet circulation
exceeds the solar age (Tassoul \cite{T00}). The slow circulation
may  affect  the internal field structure but the effect remains
small (Garaud \cite{G02}).

The global modes of the internal poloidal field of the core can
be computed by solving the eigenvalue equation,
\begin{equation}
   -\frac{A}{\tau} =
   \eta\frac{\partial^2 A}{\partial r^2}
   + \frac{\eta}{r^2}\sin\theta\frac{\partial}{\partial\theta}
   \left(\frac{1}{\sin\theta}\frac{\partial A}{\partial\theta}\right) ,
   \label{13}
\end{equation}
with the boundary condition (\ref{12}). The other condition
requires the solution to be regular at the center. We are
interested in the solutions with the longest decay times, $\tau$.

The radial profiles of the microscopic diffusivities and density
were defined after the solar structure model by Stix \& Skaley
(\cite{SS90}, see R\"udiger \& Kitchatinov \cite{RK97}). All the
computations concern  the simplest example of a stream function,
\begin{equation}
   \hat\psi \left(\theta\right) =
   -\sin\theta \bar{P}^1_2\left(\cos\theta\right),
   \label{14}
\end{equation}
where $\bar{P}^1_2$ is the normalized Legendre polynomial.
Equation~(\ref{14}) describes a flow at the base of the
convection zone from the poles to the equator, which appeared in
all our previous models for the solar differential rotation.

To describe the poloidal field geometry, the \lq escape
parameter',
\begin{equation}
 \delta\phi = \frac
 {\mathrm{max} | A\left( r,\theta\right)|_{r = R_\mathrm{in}}}
 {\mathrm{max} | A\left( r,\theta\right)|_{r \leq R_\mathrm{in}}},
 \label{15}
\end{equation}
is used. The value of $2\pi A\left( r,\theta\right)$ equals the
magnetic flux pervading a surface bounded by the longitudinal
circle of constant $r$ and $\theta$. The $\delta\phi$-parameter
(\ref{15}) estimates the ratio of the magnetic flux through the
surface of the core to the characteristic value of the flux
inside the core. The poloidal fields with small $\delta\phi$ are
expected to produce the tachocline. Linear equation (\ref{13})
leaves the amplitude of the field indefinite. The amplitude,
$B_0$, defined as the maximum strength of poloidal field within
the core, remains a free parameter of the model.

With a given poloidal field, the tachocline can be computed by
solving the equation system for the to\-ro\-i\-dal field and
angular velocity,
\begin{eqnarray}
   &&\frac{\eta}{r}
   \frac{\partial}{\partial\theta}\left(\frac{1}{\sin\theta}
   \frac{\partial\left( B\sin\theta\right)}{\partial\theta}\right)
   + \frac{\partial}{\partial r}\left(\eta
   \frac{\partial\left( Br\right)}{\partial r}\right) =
   \frac{\partial\Omega}{\partial\theta}
   \frac{\partial A}{\partial r} -
   \frac{\partial\Omega}{\partial r}
   \frac{\partial A}{\partial\theta} ,
   \nonumber \\ [.2truecm]
   &&\frac{\rho\nu}{\sin^3\theta}\frac{\partial}{\partial\theta}
   \left( \sin^3\theta\frac{\partial\Omega}{\partial\theta}\right) +
   \frac{1}{r^2}\frac{\partial}{\partial r}\left(
   r^4\rho\nu\frac{\partial\Omega}{\partial r}\right) =
   \nonumber \\ [.1truecm]
   && = \frac{1}{4\pi r^2\sin^3\theta}\left(r
   \frac{\partial A}{\partial r}
   \frac{\partial\left( B\sin\theta\right)}{\partial\theta} -
   \sin\theta\frac{\partial A}{\partial\theta}
   \frac{\partial\left( Br\right)}{\partial r}\right) .
   \label{16}
\end{eqnarray}
The neglect of the meridional flow in these equations is
justified by the small value of parameter (\ref{3}). The upper
boundary conditions  require a  zero toroidal field and the
angular velocity profile
\begin{equation}
   \Omega = 2.9 \left( 1 - 0.15\cos^2\theta\right)\ \ \
   \mu\mathrm{rad}\ \mathrm{s}^{-1}
   \label{17}
\end{equation}
favored by the helioseismology for the bottom of the convection
zone (Charbonneau et al. \cite{CDG99}).

We applied a Legendre polynomial expansion in $\theta$ to Eqs.
(\ref{13}) and (\ref{16}) and solved the resulting system of
ordinary differential equations numerically with the relaxation
method.
\section{Results and discussion}
\begin{figure}[htb]
   \centering
   \includegraphics[width=4.1cm]{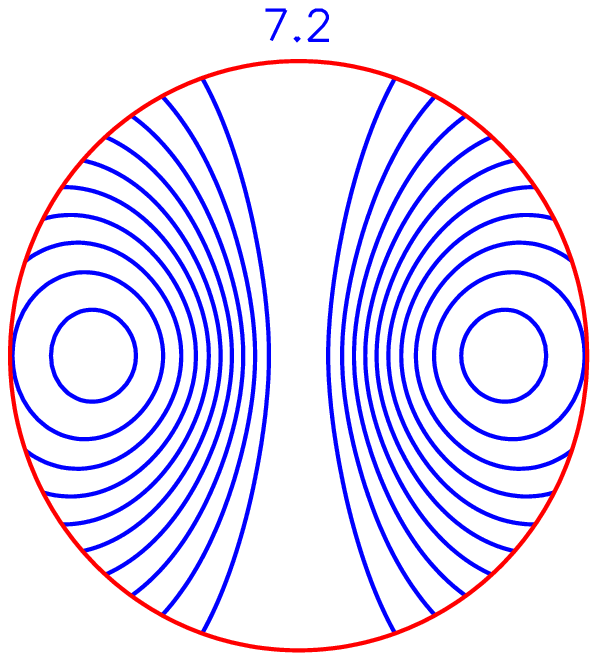}
   \hspace{0.2cm}
   \includegraphics[width=4.1cm]{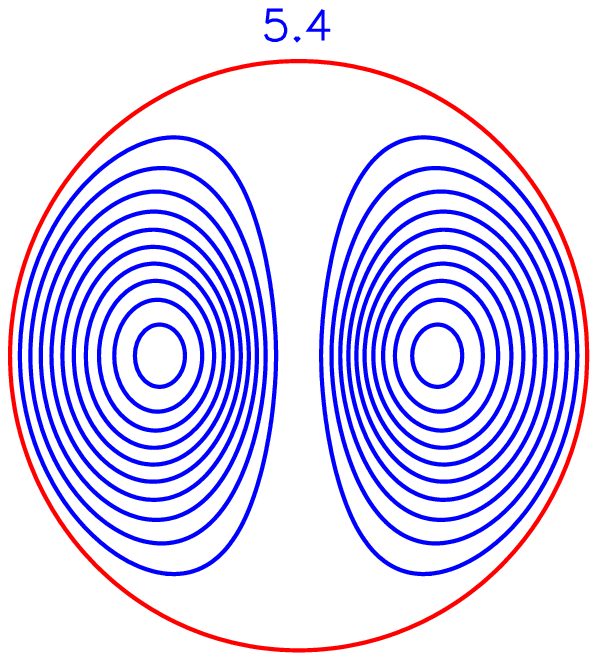}
   \caption{Field lines of the longest-living modes of poloidal field
            without  ($\rm Rm =0$,  left) and with penetration ($\rm Rm =1000$,
            right). The outer circle is the base of the convection zone. Lifetimes
            of the fields in Gyr are given at the top.
              }
   \label{f2}
\end{figure}

\begin{figure}[htb]
   \centering
   \includegraphics[width=7cm]{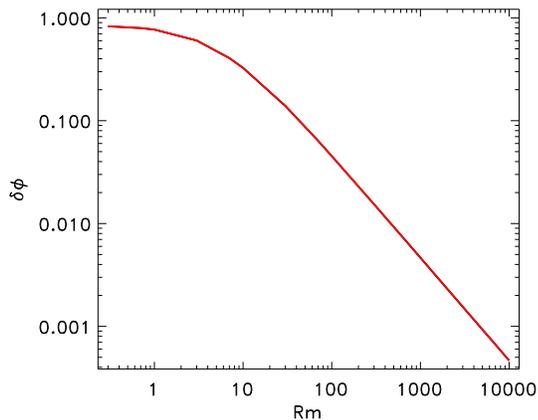}
   \caption{Geometry parameter (\ref{15}) for the longest-living dipolar
            mode of the internal field as a function of the magnetic Reynolds
            number (\ref{2}). The field becomes increasingly confined to
            the core as Rm grows.
              }
   \label{f3}
\end{figure}

Figures~\ref{f2} and \ref{f3} illustrate the effect of the
penetrating meridional flow on the structure of the longest-living
dipolar mode of an internal poloidal field. The internal field of
Fig.~\ref{f2} computed without penetration ($\rm Rm = 0$) has an
\lq open' structure. The field computed with $\rm Rm = 1000$ has
the \lq confined' geometry required for tachocline formation.
Figure~\ref{f3} shows the dependence of the escape parameter
(\ref{15}) on Rm. For $\rm Rm = 1000$, less than 1\% of magnetic
flux of the dipolar eigenmode escapes the core.

The decay time of the longest-living mode of internal field
exceeds the solar age. The lifetime decreases with Rm but
approaches a constant value of about 5.4~Gyr for large Rm. The
characteristic time of the tachocline formation, $\sim
R_\mathrm{in}/V_\mathrm{A} \sim 10^4/B_0$~yr ($V_\mathrm{A}$ is
Alfv\'{e}n velocity, $B_0$ is the field amplitude in Gauss) is
much shorter. This justifies using steady poloidal fields in the
tachocline computations.

\begin{figure}[htb]
   \centering
   \includegraphics[width=4.1cm]{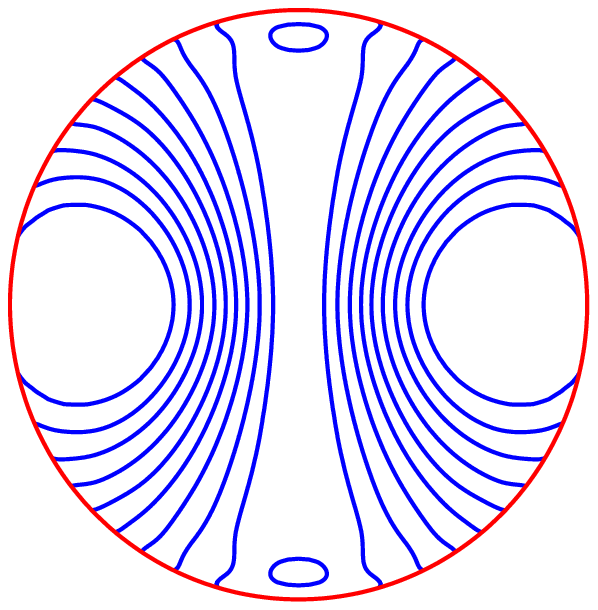}
   \hspace{0.2cm}
   \includegraphics[width=4.1cm]{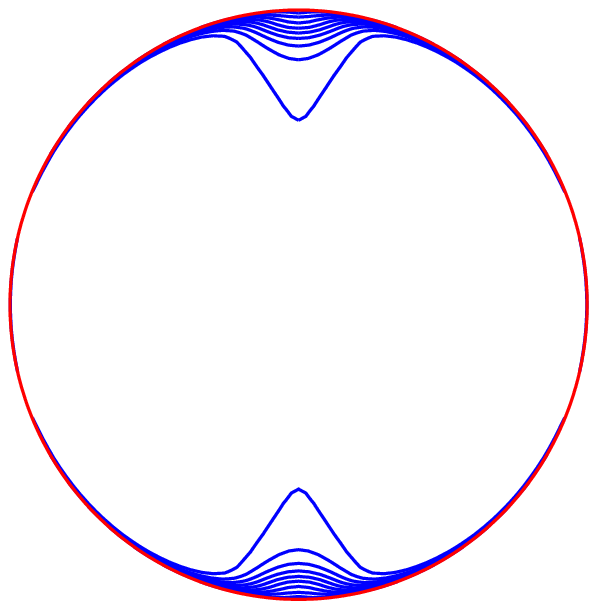}
   \caption{Angular velocity isocontours inside the radiative zone
            computed with the internal fields of Fig.~\ref{f2}.
        Left: $\rm Rm =0$, right: $\rm Rm =1000$. $B_0 = 3$~mG.
              }
   \label{f4}
\end{figure}

Figure~\ref{f4} shows the angular velocity distributions inside
the core computed for poloidal fields of Fig.~\ref{f2}. As
expected, only the confined field produces a tachocline. The
dipolar eigenmode for the solar value of ${\rm Rm}=1000$ is \lq
sufficiently confined' to do so. Already ${\rm Rm}=100$ suffices
for the tachocline formation. We conclude that the shallow
penetration of the meridional flow into the radiative core
influences the internal field geometry so strongly that it becomes
appropriate for tachocline formation. We suggest that a layer of
less than 1000 km beneath the convection zone, where a meridional
flow on the order of 1--10 m/s enters the radiative core, is
responsible for the existence of the impressive phenomenon of the
solar tachocline (which implies that the bulk of the solar core
rotates rigidly).

\begin{figure}[htb]
   \centering
   \includegraphics[width=7cm, height=5.5cm]{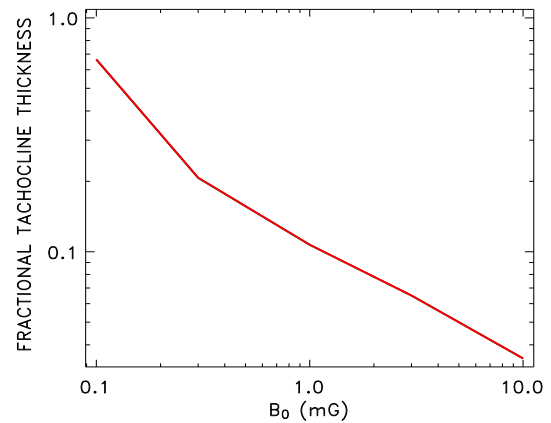}
   \caption{Dependence of the thickness of the tachocline on the
            amplitude, $B_0$, of the internal poloidal field. The thickness is
            defined as the depth of an exponential reduction in the pole-equator difference
            of the angular
        velocity.
              }
   \label{f5}
\end{figure}

Analytical estimations suggest that the (fractional) tachocline
thickness is controlled solely by the Hartmann number,
$\mathrm{Ha} =
V_\mathrm{A}R_\mathrm{in}/\sqrt{\eta_\mathrm{in}\nu_\mathrm{in}}$,
 \begin{equation}
    D_\mathrm{tach}/R_\odot \sim \mathrm{Ha}^{-1/2}
    \label{18}
 \end{equation}
(R\"udiger \& Kitchatinov \cite{RK97}). After Eq.~(\ref{18}), even
a weak internal field can produce the tachocline. The angular
velocity distributions of Fig.~\ref{f4} were computed for a field
amplitude of 3~mG. Figure~\ref{f5} shows the fractional depth
inside the core at which the angular velocity difference between
equator and poles drops e-times. When the field is not too weak,
the tachocline thickness varies as $D_\mathrm{tach}\sim
B_0^{-1/2}$ in agreement with (\ref{18}). The thickness falls
below 4\% of the solar radius at $B_0 =10^{-2}$~G. The toroidal
field in the present model is almost independent of $B_0$, and its
amplitude is not large, $B_\mathrm{tor} \sim 100$~G. The maximum
toroidal field within the core is about 140~G for ${\rm Rm}=1000$.

\begin{figure}[htb]
   \centering
   \includegraphics[width=2.8cm]{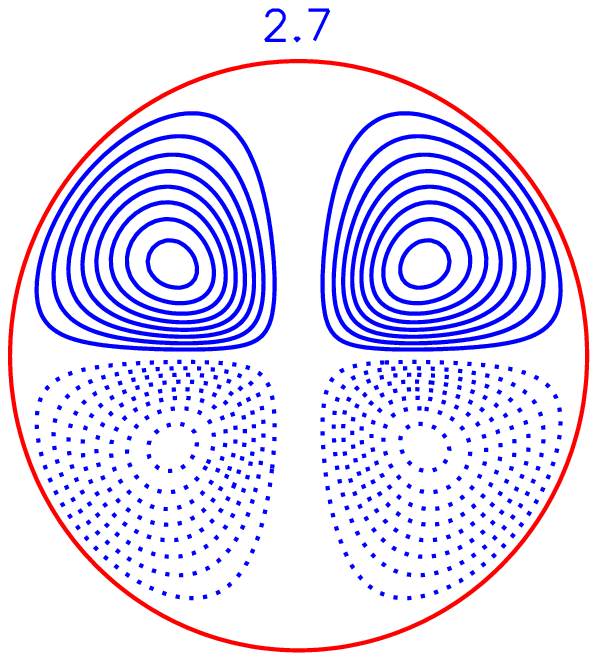}
   \includegraphics[width=2.8cm]{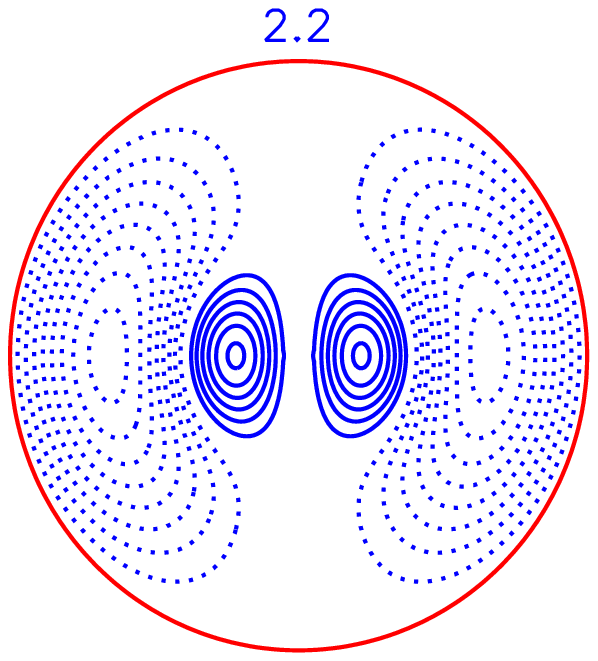}
   \includegraphics[width=2.8cm]{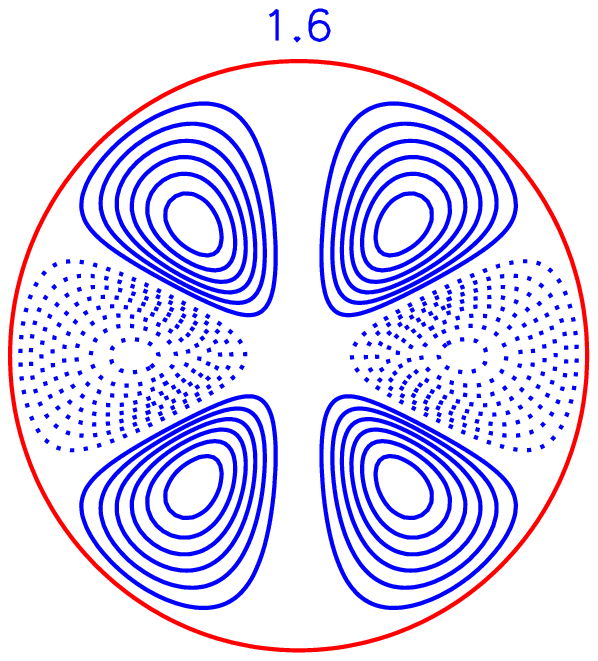}
   \caption{Eigenmodes of the internal field following the dipolar
            mode of Fig.~\ref{f2} in order of decreasing lifetime.
        The lifetimes in Gyr are given at the top. $\rm Rm =1000$.
        The geometry parameter (\ref{15}) varies as
        0.0051, 0.0073, 0.0026 from left to right.
              }
   \label{f6}
\end{figure}

The dipolar mode of Fig.~\ref{f2} has the longest lifetime. The
following modes in order of decreasing lifetime are shown in
Fig.~\ref{f6}. The lifetimes are shorter than the solar age but
not by much. The modes may be mixed in the internal field.
Tachocline computations were made for all modes of Fig.~\ref{f6},
and all the field configurations lead to a slender tachocline.

It is important for this model that the magnetic Reynolds number
(\ref{2}) be large, which is true for microscopic diffusivities or
for eddy diffusivities that are not too large, up to about
$10^8$~cm$^2$\,s$^{-1}$. The tachocline should thus be stable or
only mildly turbulent in order to allow sufficiently large Rm.

Another necessary condition is the presence of a meridional flow
of about 10~m\,s$^{-1}$ at the bottom of the  convection zone.
This flow should, of course, be present in the deep convection
zone in order to penetrate below its base. Such a flow is also a
key ingredient in the advection-dominated solar dynamo models
(Wang et al. \cite{WSN91}; Choudhuri et al. \cite{CSD95}; Dikpati
\& Gilman \cite{DG01}; Bonanno et al. \cite{Bea02}). The flow was
predicted by theoretical modeling (Kitchatinov \& R\"udiger
\cite{KR99}; Miesh et al. \cite{Mea00}), but not yet confirmed by
any other means. Probing for the deep meridional flow may be a
challenge for helioseismology.

The polar cusp in the tachocline of Fig.~\ref{f4} (right) is a
consequence of the assumed axial symmetry. This symmetry is, of
course, an idealization. We expect, though so far only on
qualitative grounds, that i) the meridional flow penetration can
also confine a nonaxisymmetric field to the core and ii) the
confined nonaxisymmetric field can also produce the characteristic
tachocline structure. An important question is still open as to
whether the axisymmetric solutions are stable against
nonaxisymmetric disturbances. Purely toroidal fields (Tayler
\cite{T73}), as well as the poloidal fields (Wright \cite{W73};
Markey \& Tayler \cite{MT73}), are known to be unstable near their
neutral lines if the fields are sufficiently strong, so that the
characteristic time of an Alfven wave passage across the star is
shorter than the rotation period. The weak fields of the present
model belong to the opposite limit.
\begin{acknowledgements}
This work was supported by the Deutsche Forschungsgemeinschaft
and by the Russian Foundation for Basic Research (project 05-02-04015).
\end{acknowledgements}

\end{document}